\documentclass{aastex}
\usepackage{emulateapj5}
\usepackage{apjfonts}

\shorttitle{SDSS Observations of GRB010222}
\shortauthors{Lee et al.}

\begin{document}
\title{Sloan Digital Sky Survey Multicolor Observations of GRB010222}

\author{
Brian C. Lee\altaffilmark{1},
Douglas L. Tucker\altaffilmark{1},
Daniel E. Vanden Berk\altaffilmark{1}, 
Brian Yanny\altaffilmark{1}, 
Daniel E. Reichart\altaffilmark{2}, 
Jennifer Adelman\altaffilmark{1},
Bing Chen\altaffilmark{3,4}, 
Mike Harvanek\altaffilmark{5},
Arne Henden\altaffilmark{6},
\v{Z}eljko Ivezi\'{c}\altaffilmark{7},
Scot Kleinman\altaffilmark{5}, 
Don Lamb\altaffilmark{8},
Dan Long\altaffilmark{5}, 
Russet McMillan\altaffilmark{5},
Peter R. Newman\altaffilmark{5}, 
Atsuko Nitta\altaffilmark{5},
Povilas Palunas\altaffilmark{9}, 
Donald P. Schneider\altaffilmark{10},
Steph Snedden\altaffilmark{5},
Don York\altaffilmark{8,11},
John W. Briggs\altaffilmark{12},
J. Brinkmann\altaffilmark{5},
Istvan Csabai\altaffilmark{3,13},
Greg S. Hennessy\altaffilmark{14},
Stephen Kent\altaffilmark{1},
Robert Lupton\altaffilmark{7},
Heidi Jo Newberg\altaffilmark{15},
Chris Stoughton\altaffilmark{1}
}

\altaffiltext{1}{Experimental Astrophysics Group, Fermi National 
Accelerator Laboratory, P.O. Box 500, Batavia, IL 60510.}

\altaffiltext{2}{Palomar Observatory, 105-24, California Institute of 
Technology, Pasadena, CA 91125.}

\altaffiltext{3}{Department of Physics and Astronomy, Johns Hopkins
University, 3701 San Martin Drive, Baltimore, MD 21218.}

\altaffiltext{4}{XMM Science Operation Center, European Space 
Agency - Vilspa, Villafranca del Castillo, Apartado 50727 - 28080 
Madrid, Spain.}

\altaffiltext{5}{Apache Point Observatory, P.O. Box 59, Sunspot, NM
88349-0059.}
 
\altaffiltext{6}{Universities Space Research Association / U. S. Naval
Observatory, Flagstaff Station, P. O. Box 1149, Flagstaff, AZ 86002-1149.}

\altaffiltext{7}{Princeton University Observatory, Peyton Hall, 
Princeton, NJ 08544-1001.}

\altaffiltext{8}{Department of Astronomy and Astrophysics, University 
of Chicago, 5640 South Ellis Avenue, Chicago, IL 60637.} 

\altaffiltext{9}{Catholic University of America, NASA/Goddard Space 
Flight Center, Code 681, Greenbelt, MD 20771.}

\altaffiltext{10}{Astronomy and Astrophysics Department, Pennsylvania 
State University, 525 Davey Laboratory, University Park, PA 16802.}

\altaffiltext{11}{Enrico Fermi Institute, University of Chicago, 
5640 South Ellis Avenue, Chicago, IL 60637.}

\altaffiltext{12}{Yerkes Observatory, University of Chicago, 
373 West Geneva Street, Williams Bay, WI 53191.}

\altaffiltext{13}{Department of Physics of Complex Systems, 
E\"otv\"os University, P\'azm\'any P\'eter s\'et\'any 1, 
H-1518, Budapest, Hungary.}

\altaffiltext{14}{U.S. Naval Observatory, 3450 Massachusetts Ave., 
NW, Washington, DC  20392-5420.}

\altaffiltext{15}{Physics Department, Rensselaer Polytechnic 
Institute, SC1C25, Troy, NY 12180.}

\begin{abstract}

The discovery of an optical counterpart to GRB010222
(detected by BeppoSAX; \citealt{gcn959}) was announced 4.4 hrs after
the burst by \citet{gcn961}.  The Sloan Digital Sky Survey's 0.5m
photometric telescope (PT) and 2.5m survey telescope were used to
observe the afterglow of GRB010222 starting 4.8 hours
after the GRB.  The 0.5m PT observed the afterglow in five, 300 sec
$g^*$ band exposures over the course of half an hour, measuring a
temporal decay rate in this short period of $F_{\nu}
\propto t^{-1.0 \pm 0.5}$.  The 2.5m camera imaged the counterpart 
nearly simultaneously in five filters ($u^* g^* r^* i^* z^*$), with
$r^*~=~18.74\pm0.02$ at 12:10 UT.  These multicolor observations,
corrected for reddening and the afterglow's temporal decay, are well
fit by the power-law $F_{\nu} \propto \nu^{-0.90 \pm 0.03}$ with the
exception of the $u^*$ band UV flux which is 20\% below this slope.  We
examine possible interpretations of this spectral shape, including
source extinction in a star forming region.

\end{abstract}


\section{Introduction}

Gamma Ray Bursts (GRBs) were first detected over three decades ago by
the Vela satellites \citep{Klebesadel73}, and the first search for
optical counterparts started nearly immediately with W. A. Wheaton's
use of the Prairie Network \citep{Grindlay74}.  These searches were
fruitless until very recently; positions accurate to a few arcminutes
were not available for days, after the bursts had decayed
substantially, placing afterglows beyond the reach of the few large
telescopes searching for them.  BATSE's near-real-time coordinates had
several degree positional errors \citep{batse} allowing only
specialized wide field instruments to respond to its triggers
\citep{Krimm95, Lee97, Akerlof99}.  The BeppoSAX satellite \citep{sax}
was the first to provide arcminute accuracy within a few hours of a
GRB.  With the early announcements of those accurate positions,
beginning in 1997, large telescopes could join the search and
discovered GRB afterglows starting with GRB970228
\citep{Groot97,vanParadijs}.

The following work describes observations of GRB010222 with the Sloan
Digital Sky Survey's (SDSS; \citealt{sdss2000}) telescopes.  The SDSS
is a project to image 10,000~$deg^2$ of the Northern Galactic Cap in
five different filters ($u^*, g^*, r^*, i^*, z^*$) to a depth of $r^*
\sim 23$ and to perform followup spectroscopy of the $10^6$ brightest
galaxies and $10^5$ quasars found in the photometry.
The SDSS is designed to be on the $u' g' r' i' z'$ photometric system
described in \citet{Fukugita} which is an $AB_\nu$ system where flat
spectrum objects ($F_\nu \propto \nu ^0$) have zero colors
\citep{Fukugita}.  The magnitudes in this paper are quoted on the 
preliminary $u^*, g^*, r^*, i^*, z^*$ system which may differ by at
most a few percent from the system of \citet{Fukugita}.
The dedicated survey instruments, a 2.5m survey
telescope and a 0.5m photometric telescope (PT), are located at Apache
Point Observatory (APO) in Sunspot, New Mexico.

\section{Observations}

\begin{figure}
\plotone{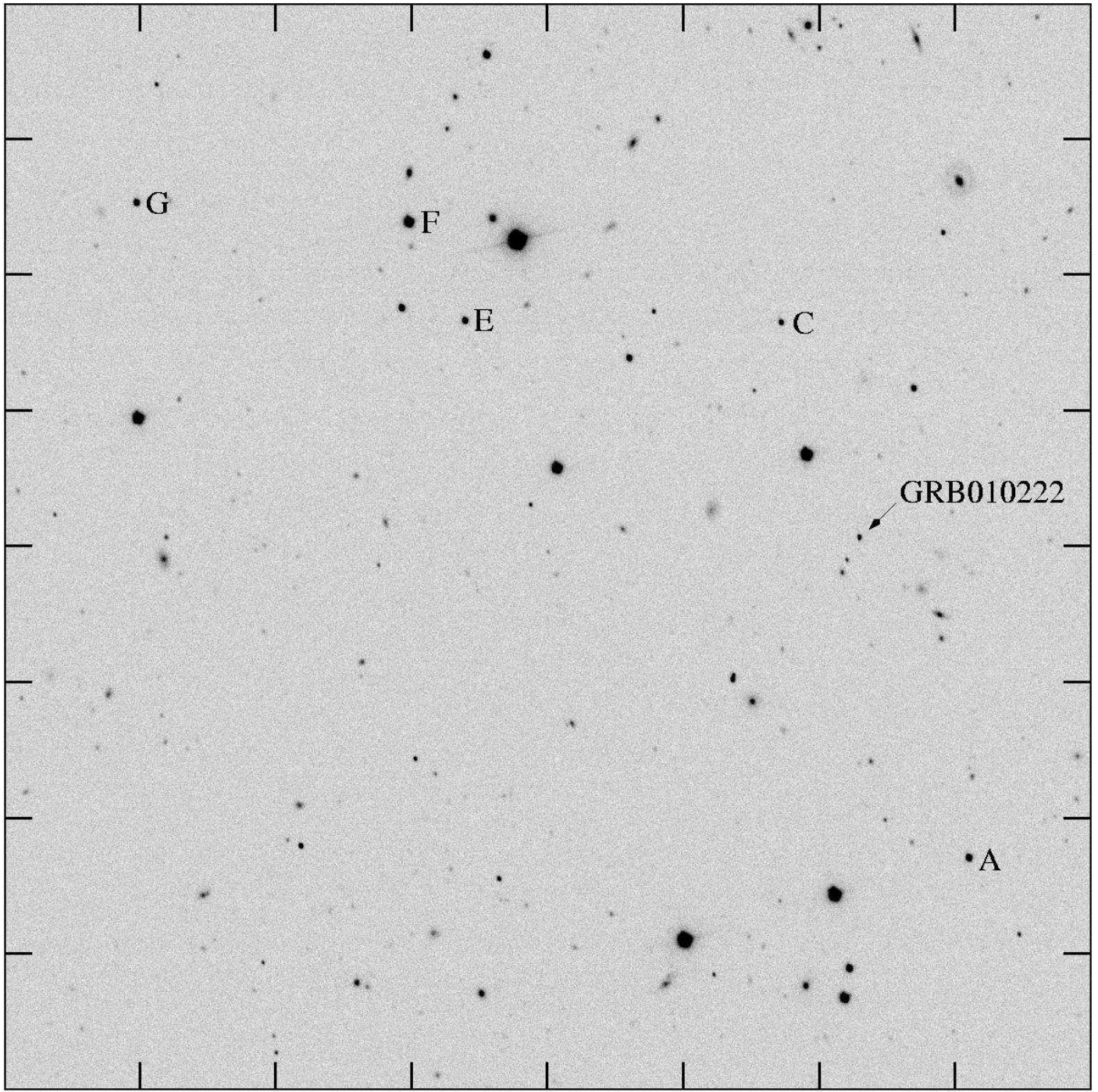}
\caption{GRB010222 2.5m telescope $r^*$ image.  The 
image is 8 arcmin square with 1 arcmin tick marks.  North is
approximately $3\degr$ clockwise from up, east is to the left.  Stars
from Table~\ref{tbl-1} within this subsection of field 22 are
indicated, including the reference star ``A'' of \citet{gcn963}.\label{fig1}}
\end{figure}

GRB010222 was detected by BeppoSAX on 
2001 February 22 at 07:23:30 U.T. in both the Gamma Ray Burst Monitor
(GRBM; 40-700 keV) and the Wide Field Camera Unit 1 (WFC1; 2-28 keV)
instruments, and was ``possibly the brightest (GRB) ever observed by
BeppoSAX'' \citep{gcn959}.  The coordinates of the BeppoSAX detection
were distributed via the GRB Coordinates Network (GCN; \citealt{gcn})
at 10:36 UT \citep{gcn959}, and \citet{gcn961} reported the discovery
of an optical counterpart at 11:48 UT, 4.4 hours after the trigger.
(See finding chart, Figure~\ref{fig1}.)  At this time conditions at
APO were not ideal for SDSS survey imaging as clouds were approaching,
and the time remaining in the night did not allow for a switch to
spectroscopy; thus SDSS observers decided to follow up the counterpart
with both the 0.5m PT and the 2.5m survey telescope.  Fortunately for
these observations the cloud passed before GRB imaging began and
conditions were more photometric after the cloud than before.

\subsection{0.5m Photometric Telescope Observations} \label{PT}

The photometric telescope is an f/8.8, 0.5m telescope equipped with
$u^* g^* r^* i^* z^*$ filters.  The single SITe $2048 \times 2048$ CCD
camera has a $41.5' \times 41.5'$ field of view.  The PT took a series
of five, 300 second observations in $g^*$ band centered on the reported
GRB010222 location, following the afterglow for
approximately 30 minutes before ending operations for the night (see
Table~\ref{tbl-1}).  Normally, the photometric telescope and the
associated reduction software are used on objects with $g^* \lesssim
18.0$.  Since the GRB exposures were unusually long and the
counterpart was relatively dim (the Poisson error limit is $\approx
3\%$) photometry was performed within a smaller than standard aperture
to improve the relative photometry of faint objects.  The counterpart
magnitudes were then corrected using a sigma clipped mean of the
magnitude offsets in each frame from the mean magnitude across the
five frames of well measured stars ($g^* \leq 17.0$); corrections were
at most 0.005 mag, indicating conditions were photometric over the 30
minute timespan.

\subsection{2.5m Survey Telescope Observations} \label{2.5m}

The 2.5m survey telescope is an f/5, 3$\degr$ field of view telescope
designed and constructed for the SDSS.  The telescope has two
interchangeable instruments, an imaging camera and a fiber-fed
spectrograph.  The imaging camera \citep{Gunn98} includes an array of
30 $2048 \times 2048$ CCDs in six columns of five CCDs each, one CCD
for each of the 5 filters.  The camera operates in a drift scan mode,
scanning the sky in great circles at sidereal rate.  Astronomical
objects are imaged for 53.9 seconds in each CCD in the order $r^* i^* u^*
z^* g^*$.  Because of the gaps between columns the telescope must
observe a second such interleaved strip to make a complete stripe.

For GRB010222 the 2.5m telescope observed two short
interleaved strips covering a roughly $2.5\degr$ square region.  The
GRB010222 counterpart was found in the second strip,
field 22 of camera column 3, run 2143.  The images were processed
through the normal SDSS data processing pipelines and calibrated
against two 0.5m PT secondary fields (hereafter, patches) of the GRB
field observed March 14 and 17.  These patches in turn were calibrated
against a system of standard stars (Smith et al., in preparation)
which the 0.5m PT observes several times throughout the night to
measure extinction and determine photometric zeropoints.  Our
diagnostic tests of the location of the stellar locus and number
counts of various classes of objects, as compared with the
approximately 1000~$deg^2$ of sky observed already in the survey,
indicate the relative calibrations are no worse than 2\% in any
filter.  The zeropoints also agree to within 1\% with a second
indirect calibration based on other 2.5m data from the same night and
four secondary patches from two previous nights.  Thus we are
confident the relative (absolute) errors are no greater than the
standard SDSS values of 3\%(5\%) for $u^*$, $g^*$, and $z^*$, and
2\%(3\%) for $r^*$ and $i^*$ (where the absolute errors include
possible differences between the SDSS preliminary photometric system
and the system of \citealt{Fukugita}).  The calibrated 2.5m magnitudes
are shown in Table~\ref{tbl-1} along with 0.5m PT observations.
Table~\ref{tbl-2} includes $u^* g^* r^* i^* z^*$ magnitudes for eleven
reference stars in field 22, including the reference star ``A'' listed
in \citet{gcn963}.  All stars have been selected from unsaturated and
non-interpolated (for cosmic ray correction) stars with $r^* \leq
19.0$ with the exception of the reference star ``A'' from
\citet{gcn963}.  This star was interpolated in $r^*$ band, but the
correction was within the quoted errors and visual inspection revealed
no problems.

\section{Results} \label{results}

\begin{figure}
\plotone{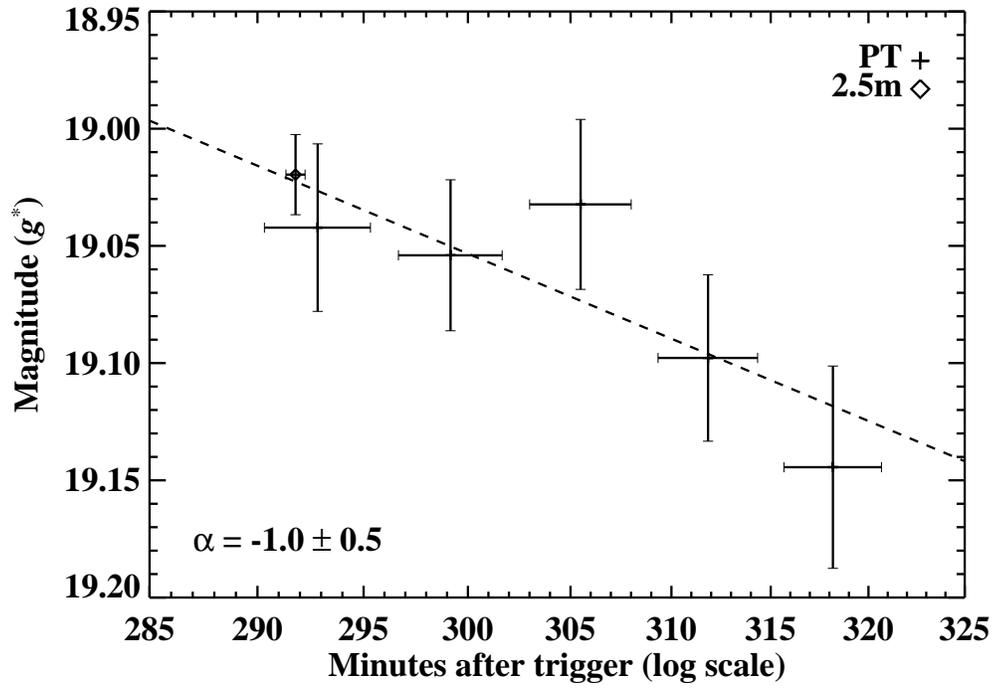}
\caption{Relative photometry for five 0.5m PT $g^*$ band observations,
along with the single 2.5m $g^*$ band observation.  The best fit decay
curve of the form $F_{\nu} \propto t^{\alpha}$ to the five PT $g^*$
band observations is $\alpha = -1.0 \pm 0.5$. \label{fig2}}
\end{figure}

The fading of GRB optical afterglows is often well fit by a power-law
decay of $F_{\nu} \propto t^{\alpha}$ with $\alpha \approx -1$
although decay rates from slightly less than this to $\alpha = -2.1$
\citep{Groot98} and breaks to steeper power-laws have been observed in
some afterglows.  GRB010222 appears to be best fit by
broken power-law models \citep{gcn1002, Masetti} with early decay
rates of $\alpha \approx -0.6$ to $-0.7$ and steeper later time decays
of $\alpha \approx -1.3$ with the break occurring around 0.5 days after
the GRB.  Considering the short time span and the limiting errors,
measurements of the decay rate with the 0.5m PT observations were
difficult.  The error weighted least squares fit to the five $g^*$
points gives $\alpha = -1.0 \pm 0.5$ (see Figure~\ref{fig2}), which is
consistent with other early decay rates measured for this burst
\citep{gcn973, gcn975, gcn983, gcn1002, Masetti}. 

The spectral shape of GRB afterglows is also well fit by a power-law,
$F_{\nu}~\propto~\nu^{\beta}$ with typical values of $\beta \approx
-1$.  In order to derive a power-law fit for our 2.5m $u^* g^* r^* i^* z^*$
observations we first corrected for the local Galactic extinction with
the dust map of \citet{schlegel} which gives extinction values of
$A_{u^*}$ = 0.118, $A_{g^*}$ = 0.087, $A_{r^*}$ = 0.063, $A_{i^*}$ =
0.048, and $A_{z^*}$ = 0.034 at the reported location $\alpha = 14^{\rm
h}52^{\rm m}12\fs55$, $\delta = +43\degr01\arcmin06\farcs2$ (J2000;
\citealt{gcn967}).  We next wished to correct for the small effect of 
fading over the few minutes between exposures in the individual bands
at 4.8 hours after the burst.  Because of the large errors in our own
decay rate measurement we instead used a least squares fit to a single
power-law for all reported $R$ band data points within 8 hours after
the burst (and before the $\approx 0.5$ day break; see
\citealt{gcn963, gcn970, gcn993, gcn1002}) with magnitudes adjusted to
the calibration of \citet{gcn987} where necessary.  The resulting fit
to the published $R$ band data is $\alpha = -0.71 \pm 0.10$,
consistent with values reported by \citet{gcn1002}.  We applied decay
corrections relative to $r^*$ of [-0.0035, -0.0066, -0.010, -0.013] to
$i^*, u^*, z^*, g^*$.

\begin{figure}
\plotone{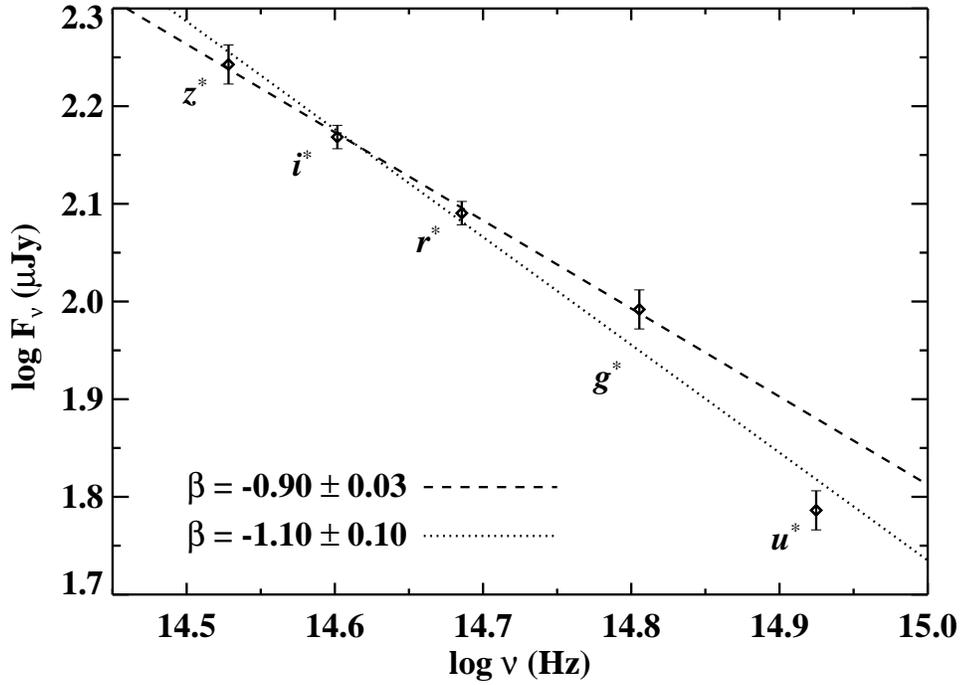}
\caption{The 2.5m multiband observations at a single epoch.  The best 
fit to $F_{\nu} \propto \nu^{\beta}$ with all five bands is $\beta =
-1.10 \pm 0.10$, shown as a dotted line.  Excluding $u^*$ band
produces a fit of $\beta = -0.90 \pm 0.03$, shown above as a dashed
line. \label{fig3}}
\end{figure}

Once these corrections were applied, the least squares fit to all five
filters is $\beta = -1.10 \pm 0.10$.  However, a much better fit can
be obtained by excluding the $u^*$ filter; the remaining non-UV filters
have a least squares fit of $\beta = -0.90 \pm 0.03$ (see
Figure~\ref{fig3}).  This second value agrees closely with the
spectral slope of $\beta = -0.89 \pm 0.03$ observed by \citet{Jha} in
a spectrum observed 4.92 hours after the burst, shortly after the 2.5m
observations.  Our $u^*$ magnitude, at an effective wavelength of
3565{\AA}, is approximately 20\% lower than the power-law fit to the
other bands.  A similar deficit was seen in the $U$ band observations of
\citet{Masetti} one and two days after the burst, and \citet{Jha} may
see the beginning of this steepening in their last binned spectra
point near 4000\AA.  (After this work was submitted similar $U$ band
results were reported by \citet{Stanek} for observations from 0.2 to
2 days after the burst.)  These independent observations indicate this
spectral feature remained constant for at least two days.

We propose that the break in the spectrum at $u^*$ may be an indication
of one of two possibilities: either the Ly$\alpha$ forest or
extinction at the source.  The first possible explanation for the $u^*$
deficit is that the counterpart is at a redshift near 2 rather than at
the redshift 1.477 absorption system reported by \citet{gcn965,
gcn974, gcn989, gcn999, Jha} and \citet{Masetti}.  For an object
without detected emission lines such as GRB010222, an absorption line
system can only provide a lower z limit to the source redshift.  The
spectrum of \citet{Jha} does not extend far past 4000{\AA}, thus no
upper limit is imposed until $z = 2.3$.  For observed QSOs the
Ly$\alpha$ forest enters $u^*$ at redshifts slightly above $z = 1.477$,
but the 20\% depression observed here would not occur unless the
redshift were $\gtrsim 2.0$ \citep{Lya}.  However \citet{Jha}
convincingly argue that given the strength of the $z = 1.477$ absorber
the GRB source is almost certainly at that redshift.  Thus a true GRB
source redshift of $\approx 2$ would seem to be an unlikely
explanation.

\begin{figure}
\plotone{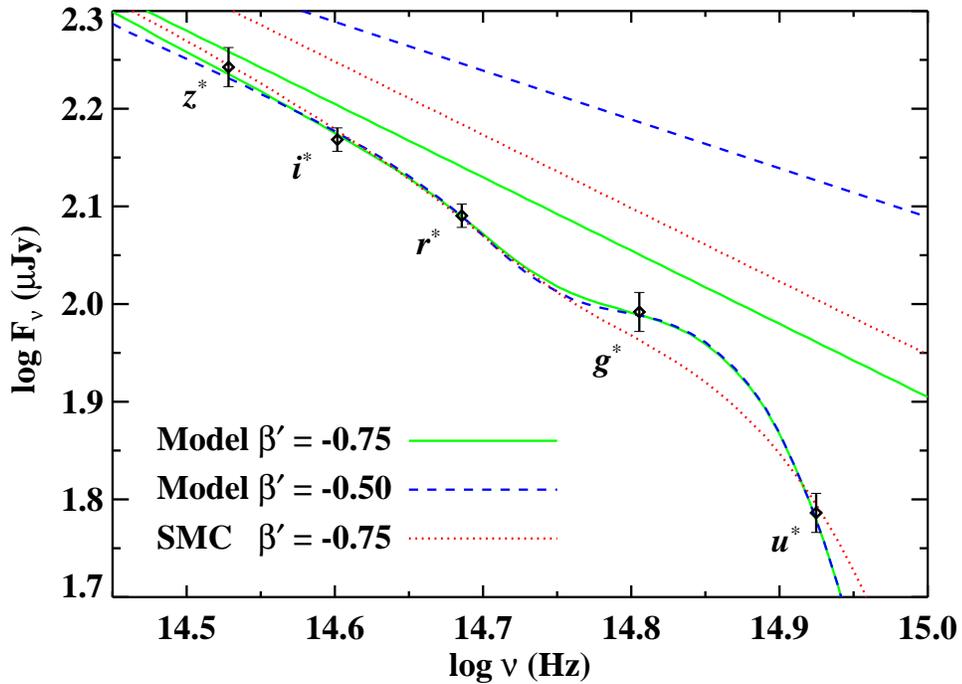}
\caption{Extinction model fits to the 2.5m multiband observations.  
The two best fit extinction models from \citet{reichart} for $\beta' =
-0.75$ and $0.5$ as discussed in section~\ref{results}, as well as an
SMC curve for an unextinguished spectrum with $\beta' = -0.75$, are
presented here.  The green and blue curves through the data points
correspond to the \citet{reichart} extinction model fit for $z =
1.477$ and a source spectrum of $\beta' = -0.75$ (green solid curve)
and $\beta' = -0.5$ (blue dashed curve).  The upper lines of the same
colors represent the corresponding unextinguished source spectra.  The
best fit \citet{reichart} parameters for $\beta' = -0.75 (-0.5)$ are a
source extinction $A_V = 0.032 (0.13)$, $c_2 = 1.35 (1.34)$, $c_3 =
8.1 (2.7)$ and $c_4 = 30 (6.9)$.  The dip in the curves, between $r^*$
and $g^*$, is the $c_3$ redshfited 2200\AA\,bump. The steep drop
through the $u^*$ observation is the $c_4$ far-UV extinction ($u^*$
samples the non-redshifted GRB spectrum from 1200\AA\,to 1600\AA).
The upper and lower dotted red lines correspond to an unextinguished
$\beta' = -0.75$ spectrum and the same spectrum extinguished by
typical SMC-like extinction with $A_V$ = 0.10. \label{fig4}}
\end{figure}

A second, more probable, explanation is that the counterpart may
reside in a star-forming region at $z = 1.477$ similar to the Large
Magellanic Cloud (LMC) or Small Magellanic Cloud (SMC) and be
extincted at the source.  Dust in front of the GRB could cause the
extinction of the afterglow and the gas would explain the large
equivalent widths in the $z = 1.477$ absorption system
\citep{York1986}.  To examine this possibility we have fit
the full extinction curve model of \citet{reichart} to the SDSS data.
Acceptable fits can be found for a wide range of intrinsic power-law
spectra (with index $\beta'$).  We present two possibilities, $\beta'
= -0.75$ and $-0.5$.  These choices for $\beta'$ are motivated as
follows: under the models of \citet{sari1998} and \citet{sari1999} the
afterglow is described in terms of synchrotron emission from a
decelerating relativistic shell or jet colliding with the surrounding
ISM.  The resulting spectrum can be expressed as four power-laws
broken at three time-dependent frequencies, the synchrotron
self-absorption frequency $\nu_a$, the cooling frequency $\nu_c$, and
the frequency corresponding to the minimum Lorentz factor of
accelerated electrons $\nu_m$.  If the shock evolves adiabatically in
a constant density medium, the break in the light curve at $\approx
0.5$ days \citep{gcn1002, Masetti} might be explained by a jet if the
observed $\nu_{opt}~>~\nu_c$ and $\nu_m$, and $\beta'
\approx -0.75$.  If the shock instead evolves radiatively, the break in
the light curve might be explained by $\nu_m$ passing through the
optical at $\approx 0.5$ days if $\nu_c < \nu_{opt} < \nu_m$, and
$\beta' \approx -0.5$.  
For $\beta' = -0.75 (-0.5)$, we find that the best fits in the
\citet{reichart} model are the source extinction 
$A_V \approx 0.032$ (0.13) mag, the slope of the UV linear component
$c_2 \approx 1.35$ (1.34), the strength of the UV bump $c_3 \approx
8.1$ (2.7), and the strength of the FUV non-linear component $c_4
\approx 30$ (6.9).  These curves are shown in Figure~\ref{fig4} 
which includes a typical SMC-like extinction curve for a source
spectrum with $\beta' = -0.75$ and source extinction $A_V$ = 0.10.
For the $\beta' = -0.75$ case there is a strong degeneracy between
$A_V$ and the parameters $c_3$ and $c_4$ such that only $A_V \cdot
c_3$ and $A_V \cdot c_4$ can be constrained; $c_3$ and $c_4$ can be
increased to any value by decreasing $A_V$, thus statistically we can
only set lower bounds.  For $\beta' = -0.75$, 
$A_V < 0.057$ mag ($1 \sigma$) and 
$A_V > 0$ at the $4.8 \sigma$ confidence level; 
$c_2 = 1.35^{+0.18}_{-0.21}$; 
$c_3 > 0$ at the $1.1 \sigma$ confidence level;
and $c_4 > 11 (1 \sigma)$, $c_4 > 0$ at the $2.5 \sigma$ confidence
level.  Further, $c_4 > 1$ (higher than any observed value) at the
$2.1 \sigma$ confidence level.
For $\beta' = -0.5$ the degeneracy is not as stong, and we can place
the following limits: $A_V = 0.13^{+0.08}_{-0.09}$ mag and 
$A_V > 0$ at the $4.8 \sigma$ confidence level; 
$c_2 = 1.34^{+0.18}_{-0.21}$;
$c_3 > 0$ at the $1.3 \sigma$ confidence level and 
$c_3 < 5.6$ ($1 \sigma$);
$c_4 > 0$ at the $2.1 \sigma$ confidence level, $c_4 > 1$ at the $1.6
\sigma$ confidence level, and $c_4 > 3.4$ ($1 \sigma$).
The \citet{reichart} fits to the two GRB models are approximately
equally likely (with the $\beta' = -0.75$ model fit only 1.4 times
more probable than the $\beta' = -0.5$ model fit).  

The \citet{reichart} best fit values of $c_2$ and the second value of
$c_3$ are typical of that observed in the LMC.  Given the errors, the
first value of $c_3$ is not inconsistent with this interpretation.
However, for both afterglow models the values of $c_4$, required to
extinguish $u^*$ relative to the other bands, are about an order of
magnitude greater than those found in the LMC or SMC.  \citet{Waxman}
and \citet{Galama} propose that the optical flash (e.g.,
\citealt{Akerlof99}) and the burst may sublimate and fragment dust in
the circumburst environment.  If small (radius $<$ 300 $\rm{\AA}$)
graphite grains, which may be responsible for the FUV non-linear
component \citep{Draine1984}, survive in greater numbers in this
environment, this value of $c_4$ is not unreasonable.  Alternatively,
the large $c_4$ value and $u^*$ band deficit could be due to
absorption by molecular hydrogen
\citep{Draine2000} which would span the entire $u^*$ band at $z =
1.477$; however the expected feature at $\lambda \leq 1650$\AA\,
redshifted to $\lambda \approx 4000$\AA\, is not obvious in the
published spectra.

\section{Conclusions}

The serendipitous 2.5m survey telescope observations of
GRB010222 occurred in this case because of the very fortunate
timing of a counterpart discovery announcement towards the end of a
night when conditions did not favor normal survey operations.  The
2.5m camera is an unwieldy instrument for rapid followup observations;
nonetheless this observation has shown the value of early five filter
observations.  In addition to the measurement of the $g^* r^* i^* z^*$
spectral slope ($F_{\nu}~\propto~\nu^{-0.90\pm0.03}$), the break to a
steeper slope in $u^*$ (also seen in the $U$ band observations of
\citealt{Masetti} and \citealt{Stanek}) was not predicted or seen in 
spectra, and may indicate an alternate source redshift, source
extinction in a star forming region modified by the GRB or its
progenitor, or something else entirely.

The 0.5m PT is an automated telescope and in general much better
suited for GRB followup observations than the 2.5m survey telescope.
In this case the same timing that was so fortunate for the 2.5m was
disadvantageous for the PT, which was only able to observe the burst
near its limit and for a short period before shutting down for the
night.  Due to an afterglow's rapid decay, typical BeppoSAX delays of
several hours place afterglows near the detection limit of smaller
telescopes.  HETE-2, launched in October of 2000, will soon provide
$\sim 10$ arcminute positions for GRBs within minutes of the trigger,
potentially allowing telescopes such as the PT to measure both the
spectral and temporal behavior of a burst in the first few hours.

\acknowledgments
\section*{Acknowledgments}

We would like to thank Scott Barthelmy and everyone who has made the
GCN possible, as well as the BeppoSAX team who provided the
localization so crucial to this work.  We would also like to thank
Bruce Woodgate for his help in these observations.

The Sloan Digital Sky Survey (SDSS) is a joint project of The
University of Chicago, Fermilab, the Institute for Advanced Study, the
Japan Participation Group, The Johns Hopkins University, the
Max-Planck-Institute for Astronomy (MPIA), the Max-Planck-Institute
for Astrophysics (MPA), New Mexico State University, Princeton
University, the United States Naval Observatory, and the University of
Washington. Apache Point Observatory, site of the SDSS telescopes, is
operated by the Astrophysical Research Consortium (ARC).

Funding for the project has been provided by the Alfred P. Sloan
Foundation, the SDSS member institutions, the National Aeronautics and
Space Administration, the National Science Foundation, the
U.S. Department of Energy, Monbusho, and the Max Planck Society. The
SDSS Web site is http://www.sdss.org/.

\clearpage

\begin{deluxetable}{ccccr}
\tablecaption{SDSS Observations of the GRB010222 Afterglow.\label{tbl-1}}
\tablewidth{0pt}
\tablehead{
\colhead{UTC\tablenotemark{a}} & \colhead{telescope} & \colhead{band}
& \colhead{exposure (sec)}
& \colhead{magnitude\tablenotemark{b}}
}
\startdata
12:09:35 &2.5m &$r^*$ &54 &$18.74\pm0.02$\\
12:10:47 &2.5m &$i^*$ &54 &$18.53\pm0.02$\\
12:11:59 &2.5m &$u^*$ &54 &$19.56\pm0.03$\\
12:13:10 &2.5m &$z^*$ &54 &$18.34\pm0.03$\\
12:14:22 &2.5m &$g^*$ &54 &$19.02\pm0.02$\\

12:13:15 &0.5m &$g^*$ &300 &$19.04\pm0.04$\\
12:19:36 &0.5m &$g^*$ &300 &$19.05\pm0.04$\\
12:25:58 &0.5m &$g^*$ &300 &$19.03\pm0.04$\\
12:32:19 &0.5m &$g^*$ &300 &$19.10\pm0.04$\\
12:38:41 &0.5m &$g^*$ &300 &$19.14\pm0.04$\\
\enddata

\tablenotetext{a}{Exposure start time, 2001 Feb 22.}

\tablenotetext{b}{Statistical errors; absolute photometry errors for
the 2.5m may be as large as 5\% for $u^*$, $g^*$, and $z^*$, 3\% for $r^*$
and $i^*$.}

\end{deluxetable}

\begin{deluxetable}{cccccccc}
\tabletypesize{\scriptsize}
\tablecaption{Reference Stars in the GRB010222 Field.\tablenotemark{a}\label{tbl-2}}
\tablewidth{0pt}
\tablehead{
\colhead{} & \colhead{$\alpha$\tablenotemark{b}} & \colhead{$\delta$\tablenotemark{b}} & \colhead{$u^*$} & \colhead{$g^*$}& \colhead{$r^*$}& \colhead{$i^*$}& \colhead{$z^*$}}
\startdata
GRB & $14^{\rm h}52^{\rm m}12\fs51$ & $+43\degr01\arcmin06\farcs2$ & $19.56\pm0.03$ & $19.02\pm0.02$ & $18.74\pm0.02$ & $18.53\pm0.02$ & $18.34\pm0.03$\\
A   & $14^{\rm h}52^{\rm m}07\fs51$ & $+42\degr58\arcmin48\farcs6$ & $19.39\pm0.02$ & $17.96\pm0.02$ & $17.40\pm0.02$ & $17.18\pm0.01$ & $17.08\pm0.01$\\
B   & $14^{\rm h}52^{\rm m}12\fs57$ & $+42\degr55\arcmin59\farcs3$ & $18.29\pm0.02$ & $17.02\pm0.02$ & $16.48\pm0.02$ & $16.26\pm0.01$ & $16.10\pm0.01$\\
C   & $14^{\rm h}52^{\rm m}16\fs06$ & $+43\degr02\arcmin38\farcs5$ & $20.02\pm0.03$ & $19.03\pm0.02$ & $18.72\pm0.02$ & $18.61\pm0.01$ & $18.56\pm0.03$\\
D   & $14^{\rm h}52^{\rm m}21\fs86$ & $+42\degr56\arcmin29\farcs0$ & $19.78\pm0.03$ & $18.42\pm0.02$ & $17.87\pm0.02$ & $17.67\pm0.01$ & $17.53\pm0.02$\\
E   & $14^{\rm h}52^{\rm m}28\fs65$ & $+43\degr02\arcmin32\farcs6$ & $20.37\pm0.04$ & $18.50\pm0.02$ & $17.75\pm0.02$ & $17.45\pm0.01$ & $17.29\pm0.02$\\
F   & $14^{\rm h}52^{\rm m}31\fs09$ & $+43\degr03\arcmin14\farcs2$ & $17.06\pm0.01$ & $15.89\pm0.02$ & $15.48\pm0.01$ & $15.30\pm0.01$ & $15.26\pm0.01$\\
G   & $14^{\rm h}52^{\rm m}41\fs98$ & $+43\degr03\arcmin16\farcs8$ & $19.37\pm0.02$ & $18.15\pm0.02$ & $17.69\pm0.02$ & $17.52\pm0.01$ & $17.42\pm0.02$\\
H   & $14^{\rm h}52^{\rm m}45\fs90$ & $+42\degr57\arcmin09\farcs2$ & $20.07\pm0.04$ & $17.80\pm0.02$ & $16.81\pm0.02$ & $16.46\pm0.01$ & $16.24\pm0.01$\\
I   & $14^{\rm h}52^{\rm m}50\fs13$ & $+42\degr55\arcmin22\farcs7$ & $17.84\pm0.01$ & $16.51\pm0.02$ & $15.99\pm0.01$ & $15.83\pm0.01$ & $15.73\pm0.01$\\
J   & $14^{\rm h}52^{\rm m}51\fs32$ & $+42\degr54\arcmin56\farcs3$ & $18.53\pm0.02$ & $16.49\pm0.02$ & $15.63\pm0.01$ & $15.37\pm0.01$ & $15.18\pm0.01$\\
K   & $14^{\rm h}52^{\rm m}51\fs85$ & $+43\degr03\arcmin05\farcs7$ & $18.59\pm0.02$ & $16.91\pm0.02$ & $16.28\pm0.01$ & $16.01\pm0.01$ & $15.89\pm0.01$\\
\enddata

\tablenotetext{a}{Selected from stars in field 22 with magnitude 
$r^* \leq 19.0$ which were well measured, unsaturated, and
non-interpolated (cosmic ray corrected) in all five filters, with the
exception of the first star (A) in the list which is the reference
star ``A'' listed in \citet{gcn963}.  This star was interpolated in
$r^*$ band but manual inspection revealed no problems; the correction
was small and within the quoted errors.  Listed errors are statistical
only, absolute photometry errors for the 2.5m may be as large as 5\%
for $u^*$, $g^*$, and $z^*$, 3\% for $r^*$ and $i^*$.}

\tablenotetext{b}{Because of the non-standard orientation and short
length of this stripe the astrometric errors are unusually large,
approximately $0\farcs3$.}
\end{deluxetable}

\clearpage

\epsscale{1.0}

\epsscale{0.75}

\clearpage

\end{document}